\documentclass[epjST]{svjour}
\usepackage{graphics}
\usepackage{amssymb}
\begin{document}
\title{Ground state cooling of nanomechanical resonators by electron transport}
\subtitle{ }
\author{G. Rastelli \inst{1,2}\fnmsep\thanks{\email{gianluca.rastelli@uni-konstanz.de}} \and W. Belzig\inst{2} }
\institute{Zukunftskolleg,Universit{\"a}t Konstanz, D-78457 Konstanz, Germany \and Fachbereich Physik, Universit{\"a}t Konstanz, D-78457 Konstanz, Germany}
\abstract{
We discuss two theoretical proposals for controlling the nonequilibrium
steady state of nanomechanical resonators using quantum
electronic transport. 
%s
Specifically, we analyse two approaches to achieve the ground-state cooling of the mechanical vibration coupled to a quantum dot embedded  
between (i)  spin-polarised contacts or (ii)  a normal metal and a superconducting contact. 
Assuming a suitable coupling between the vibrational modes and the charge or spin of the electrons in the quantum 
dot, we show that ground-state cooling of the mechanical oscillator 
is within the state of the art  for suspended carbon nanotube quantum dots operating as electromechanical   devices.
} %end of abstract
\maketitle
\section{Introduction}
\label{intro}
Mesoscopic conductors coupled to localised, quantum harmonic resonators have now become a commonly studied system, 
both experimentally and theoretically. 
Interesting phenomena in such systems arise from the interplay between 
the resonator dynamics and the quantum transport in the single electron regime. 
The localised oscillator modes can be either a microwave photon cavity \cite{Childress:2004,Viennot:2015,Mi:2017} 
or a mechanical resonator  \cite{Poot:2012}.
In the latter case, these systems operate as electromechanical systems and they 
include suspended carbon nanotube quantum dots \cite{Benyamini:2014,Weber:2014}, 
quantum dots in suspended semiconductor membranes  \cite{Weig:2004},
quantum dots coupled to a piezoelectric nanoresonator \cite{Okazaki:2016}, 
or superconducting single-electron transistors \cite{Blencowe:2005}. 
Such electromechanical systems typically operate far from equilibrium and can be very strongly nonlinear, 
allowing us to unveil quantum dynamical properties unexplored so far.
They are  also interesting to address fundamental issues as they are expected to enter the quantum regime at low temperature 
and hence  open the route for fundamental tests of quantum mechanics in massive objects \cite{Blencowe:2004}.

Suspendend carbon nanotube quantum dots (CNT-QD) are a priori good candidates for realising {\sl quantum} 
electromechanical systems: 
(i) mechanical modes can reach extremely high quality factors $Q\sim10^{6}$ without detriment 
of the electron transport properties  \cite{Moser:2014}; 
(ii) recent experiments  showed unprecedented control of the tunability of both electron transport and 
electromechanical interaction \cite{Benyamini:2014}.
To achieve the quantum regime of the mechanical motion, a crucial requirement is cooling the system to a temperature much lower 
than the characteristic frequency, viz. $k_BT \ll h f$ with $h$ the Planck's constant and $f$ the frequency of a mechanical mode.
In this way, starting from the ground state, one aims to have access and control of only few low energy excitations of the quantum oscillator.
Despite some progress, this goal still remain to be achieved in the flexural mechanical modes of suspended CNTs.
The crucial problem is the low frequency of the flexural modes whose typical value is around hundreds of MHz, or below.
This implies that the electromechanical devices would have to be cooled to extremely cryogenic temperature below few milli-kelvin which is a demanding task 
in the low temperature electronic circuitry.

%%%%%%%%%%%%%%%%%%%%%%%%%%%%%%%%%%%%%%%%%%%%%%%%%%%%%%
%
%			FIGURE N.1
%
%
%%%%%%%%%%%%%%%%%%%%%%%%%%%%%%%%%%%%%%%%%%%%%%%%%%%%%%
%
%
\begin{figure}[b]
%\resizebox{0.75\columnwidth}{!}{%
%
\resizebox{1.\columnwidth}{!}{ 
\includegraphics{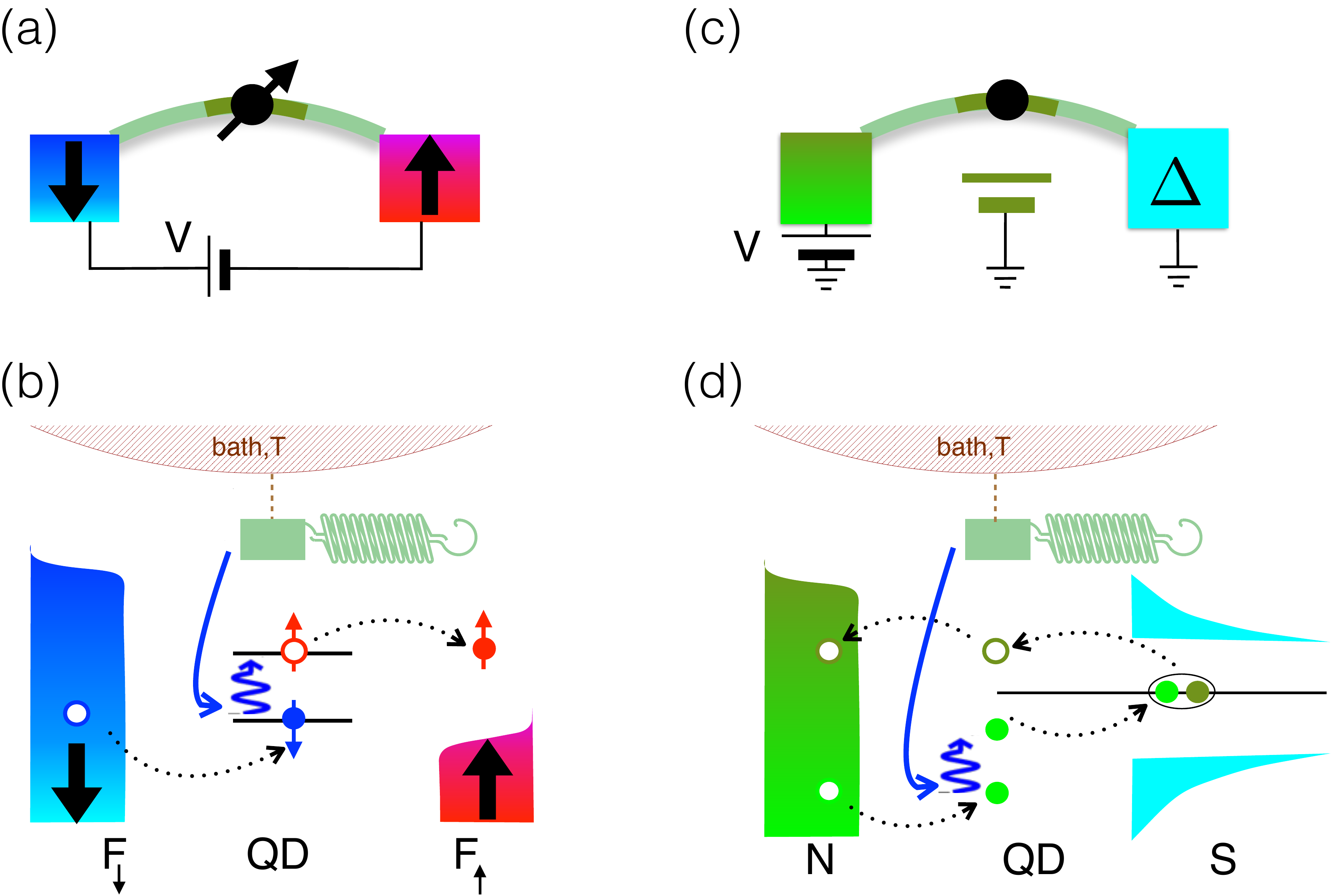}
}
\caption{A quantum dot (QD) is formed when electrons are confined to a small region within a carbon nanotube (CNT) suspended between 
two conducting leads. 
(a) The spin of the QD electron states is coupled to the flexural modes of the CNT  suspended between 
two ferromagnetic contacts of opposite polarization (see text).
(b) Schematic picture of the microscopic model for (a). The QD corresponds to two spin levels with a Zeeman splitting and a single flexural mode to an oscillator.
The spin-vibration interaction leads to vibration-assisted inelastic spin-flip processes accompanied by the exchange (e.g. absorption) of energy 
with the oscillator.
(c) The charge of  QD electron states is coupled to the flexural modes of the suspended between a normal metal N and a superconductor S 
with gap $\Delta$ (see text).
(d) Schematic picture of the microscopic model for (c). 
The QD  corresponds to two spin-degenerate levels and a single flexural mode to an oscillator.
At small bias voltage $V$, the charge-vibration interaction leads to vibration-assisted inelastic Andreev reflections accompanied 
by the exchange (e.g. absorption) of energy  with the oscillator.
}
\label{fig:1} 
\end{figure}
%
%
%
%%%%%%%%%%%%%%%%%%%%%%%%%%%%%%%%%%%%%%%%%%%%%%%%%%%%%%

Several and interesting theoretical proposals have been analyzed for achieving active cooling by  using electron 
transport \cite{Martin:2004,Brown:2007,Pistolesi:2009,Zippilli:2009,Santandrea:2011,Sonne:2011}, 
exploiting  the effect of the back-action force on the oscillator due to the interaction with a 
mesoscopic conductor. 
Most of them are closely related to the optical mechanism of the side-band cooling \cite{Aspelmeyer:2014} which, in a scattering picture, 
is based on the enhancement of phonon absorption due to the matching of the oscillator's frequency $f$  
with a resonant excitation of the conductor.

In this paper we discuss two proposals for cooling flexural modes of a suspended CNT-QD 
using electron transport.

The first system is a nanomechanical spin-valve.
The injection of spin-polarized current has been experimentally
reported in CNT-QDs in a spin-valve geometry with gate field control 
and with ferromagnetic nanocontacts \cite{Sahoo:2005}.  
Moreover the spin of discrete electron levels in the dot is theoretically predicted to couple to the 
flexural vibrations due to the mediation of the intrinsic spin-orbit interaction \cite{Palyi:2012} or due to the presence 
of an external magnetic gradient \cite{Atalaya:2012}. 
We combine these two aspects and propose the system sketched  in Fig.\ref{fig:1}(a,b), \cite{Stadler:2014,Stadler:2015}.

The second system is a quantum dot connected to one superconducting contact and a normal metal as shown in Fig.\ref{fig:1}(c,d).
In this system, for energy scales involved in the transport and smaller than the superconducting gap $\Delta$,
finite current flows through the system due to Andreev Reflection (AR) in which, for instance, an incoming electron 
from the normal lead is reflected as hole with the concurrent formation of a Cooper pair into the superconductor.
In the presence of an interaction of the quantum dot with bosonic modes of the environment, 
AR can be inelastic  and 
experimental observations of such inelastic reflections in CNT-QD have been reported \cite{Gramich:2015}.
We consider a microscopic model of charge-vibration interaction in the QD 
with phonon emission or absorption in the vibration-assisted Andreev Reflection \cite{Stadler:2016,Stadler:2017}.

This paper is organized as follows.
In the section \ref{sec:2} we start with a general, theoretical approach to discuss the electromechanical effects in quantum dots 
coupled to local resonators.  
We discuss how the nonsymmetrised noise of the dot's operator coupled to the vibration determines two important 
electromechanical effects: the induced damping and the steady, nonequilibrium  phonon occupation.
In the following section we analyze the behavior of these two quantities.
In the section \ref{sec:3}  we report the results for the  first model shown in Fig.1(a,b) whereas the section \ref{sec:4} 
contains the results for the second model shown in Fig.1(c,d). 
Beyond the phase diagram of the phonon occupation in terms of the bias voltage and of the gate voltage,
we explain how information about the resonator's non-equilibrium state can be extracted by 
distinct features of the inelastic current. 
In the section \ref{sec:5} we summarise our conclusions.

%%%%%%%%%%%%%%%%%%%%%%%%%%%%%%%%%%%%%%%%%%%%%%%%
%
%
%
\section{Electromechanical model}
\label{sec:2}
Quantum dots in real devices can be modeled as a single-impurity Holstein model in which 
one assumes a linear coupling between the electron occupation on the quantum dot and the 
oscillation amplitude of one (or more harmonic modes) representing the local 
vibrations \cite{Galperin:2004}. 
Here we generalise this model and consider the following model Hamiltonian
\begin{equation}
\hat{H} = \sum_{\alpha=l,r} \left(  \hat{H}_{\alpha} + \hat{H}_{\alpha,t} \right) 
+ \sum_{\sigma=\uparrow,\downarrow} \varepsilon_{\sigma} \hat{n}_{\sigma}
+ \lambda \, \hat{F}_d \,  \, ( \hat{b}^{\phantom{g}} + \hat{b}^{\dagger} )
+ \omega_0  \, \hat{b}^{\dagger}  \hat{b}^{\phantom{g}}  \, ,
\label{eq:H_start}
\end{equation}
where $\hat{H}_{\alpha} $ is the Hamiltonian for the left and right lead $(\alpha=l,r)$, $ \hat{H}_{\alpha,t}$ is the tunneling 
Hamiltonian between the dot and the leads (we set $\hbar=1$).
The nature of these contacts will be specified in the next two sections, for two different cases.
The operator $\hat{b}$ and $\hat{b}^{\dagger}$ are the (bosonic) creation and annihilation operators of the harmonic oscillator of frequency $\omega_0$
and $\hat{d}_{\sigma}$ and $\hat{d}^{\dagger}_{\sigma}$ are the corresponding fermionic operators for the dot's levels.
The coupling strength of the interaction is $\lambda$.
The operator $\hat{F}_d$ is the force acting on the oscillator.
We will study the case when  $\hat{F}_d$ corresponds to the $x$ component of the local spin operator
$\hat{F}_d = \hat{s}_x = \hat{d}^{\dagger}_{\uparrow}  \hat{d}^{\phantom{g}}_{\downarrow} 
+  \hat{d}^{\dagger}_{\downarrow}  \hat{d}^{\phantom{g}}_{\uparrow} $ in section \ref{sec:3},
whereas in section \ref{sec:4} we analyse the case when $\hat{F}_d$ corresponds to the total
charge.   

Furthermore, we assume the weak coupling limit regime given by 
\begin{equation}
\lambda  \ll  \omega_0 \, .
\end{equation}
This means that the variation of the charge or the spin in the dot induces a displacement of the energy of the order of $\lambda$ 
which is small compared to the level separation of the harmonic oscillator. 
In that case, polaronic effects are negligible and the bare levels and states of the harmonic oscillator are meaningful starting points 
to deal with in presence of the electron-vibration interaction and current flowing through the dot.
We aim to focus on sharp resonance transport regime so that we also require another condition for the typical tunneling rate $\Gamma$ 
controlling the hopping of the electrons from the leads to the dot
\begin{equation}
\label{eq:lambda_gamma}
\Gamma \ll  \omega_0 \, . 
\end{equation}
Since the inverse of the tunneling rate $\hbar/\Gamma$ is related to the dwell time of the electron in the dot,  
this condition is known as the anti-adiabatic regime, in which the fast oscillator readjust to the variations of the charge 
or spin in the dot due to the quantum tunneling.
Assuming the weak coupling and anti-adiabatic regime, we calculate two important quantities:
the nonequilibrium occupancy of the harmonic
oscillator $\bar{n}_c$ and the inelastic current through to the dot $I_{in}$ due to the electron-
vibration interaction in the leading order of $\lambda^2$.

\subsection{Electromechanical damping}
When a voltage bias is applied, the  electrons tunneling through quantum dot behave as an effective environment 
characterized by an electromechanical damping $\gamma$ and a force noise acting on the oscillator.
Then the crucial quantity is the unperturbed, non-symmetrized noise of the electron force operator (charge or spin) of the dot
\begin{equation}
\label{eq:S_w_force}
S(\omega) 
=   \,  \int^{+\infty}_{\infty}\!\!\!\!\!\!\!\! dt  \,\, e^{i \omega t} \, {\langle  \hat{F}_d(t) \hat{F}_d(0)  \rangle}_{\lambda=0}
%= \left\{
%\begin{array}{c}
% S_{+}(\omega) \quad \mbox{for} \quad \omega >0  \\
% S_{-}(\omega) \quad \mbox{for} \quad \omega<0   
%\end{array}
%\right.
. 
\end{equation}
with ${\langle \dots \rangle}_{\lambda=0}$ denoting the quantum statistical average taken over 
the electron system for $\lambda=0$.
Then, we can express the electromechanical damping as
\begin{equation}
\gamma =  \lambda^2 \left[ S(\omega_0) -  S(-\omega_0) \right] \equiv  
 \gamma_{+}  -   \gamma_{-}
 \, .
\end{equation}
In other words, the absorption of an energy quantum $\omega_0$ is connected to the intrinsic
non-symmetrized noise at the positive frequency of the open dot (non interacting with
the vibration) whereas the emissions of an energy quantum $\omega_0$ is connected to the
non-symmetrized noise at the negative frequency.
A simple way  to understand the relation between the non-symmetrized noise $\gamma_{\pm}$
and the probability of absorption or emission  of a phonon of energy $\hbar \omega_0$ is based on Fermi's Golden rule.
For the probability per unit time of one phonon absorption $(+)$ or  emission $(-)$, the Golden rule gives
\begin{equation}
p_{\pm} =
2\pi \sum_{n} \sum_{i,f} \, P_{n} \, P_i \, 
{\left| 
\langle n \mp 1,\psi_f  \right| \hat{H}_{int}   \left| n,  \psi_i \rangle 
\right|}^2 
\delta\left[   \omega_0  \pm \left( E_i -E_f \right) \right]  \, ,
\end{equation}
where the $\psi_{i}$ and $\psi_{f}$  are the initial and final states of the open dot,  
with energies $E_i$ and $E_f$, and $P_i$ is the probability of occupation of the initial state 
whereas $P_n$ is the probability of occupation of the Fock state $\left| n  \right>$.
Using the integral representation for the $\delta-$function and 
the interaction $ \hat{H}_{int}  = \lambda \hat{F}_d ( \hat{b} + \hat{b}^{\dagger}   ) $, 
one obtains for the case of absorption 
\begin{eqnarray}
p_{+}  & =& 
\lambda^2 \sum_{n}  n P_n 
\sum_{i,f}  \, P_i \, 
\left< \psi_i \right| \hat{F}_d   \left| \psi_f  \right>   
\left< \psi_f \right| \hat{F}_d   \left| \psi_i  \right>   
\, 
\int^{+\infty}_{-\infty} \!\!\!\!  dt \,e^{i  \left[ \omega_0 +   E_i -E_f   \right] t} \\
& =&
\lambda^2 \bar{n}
\int^{+\infty}_{-\infty}  \!\!\!\! dt \, e^{i  \omega_0 t} 
{\left< \hat{F}_d(t) \hat{F}_d (0)
\right>}_{\lambda=0} =  \gamma_{+}  \,\, \bar{n} \, ,
\end{eqnarray} 
in which the completeness of the dot's states was used  and  we set $\bar{n} = \sum_n n P_n$.
A similar calculation for the emission of one phonon lead to 
\begin{equation}
p_{-} =   \left(\bar{n}+1  \right)  \,   \gamma_{-}  \, .
\end{equation}

\subsection{Nonequilibrium steady state}
In order to calculate the steady state nonequilibrium occupation $n$ due to the charge-vibration interaction, 
we neglect in a first approximation the thermal bath and use an heurestic and phenomenological approach by assuming following equation rate 
\begin{equation}
\label{eq:phenom}
0 = \frac{d  \bar{n} }{dt} =
\bar{n}  \gamma_{+}  -  \left( \bar{n} + 1   \right)    \gamma_{-} 
\longrightarrow  
\bar{n}  =  \frac{ \gamma_{-}}{ \gamma_{+}- \gamma_{-}} \equiv \bar{n}_c \, .
\end{equation}
The result for $n$ Eq.~(\ref{eq:phenom}) clearly points out that ground state cooling with $\bar{n}_c \ll 1$  
can be reached for  $\gamma_{+} \gg \gamma_{-}$.
In other words,  one needs to create a strong asymmetry between the two processes in order 
to cool the oscillator.
Hereafter, we call the coefficients  $\gamma_{\pm}$ the {\sl intrinsic} rates 
or simply {\sl rates for the phonon emission and absorption} since they are a property 
of the intrinsic system without the interaction with the resonator.

One can generalize Eq.~(\ref{eq:phenom}) taking into account the (unavoidable) 
interaction of the oscillator with a thermal bath with an intrinsic damping rate $\gamma_0$.
Then, the general steady occupation of the oscillator is given by the competition 
between the interaction of the oscillator with the effective environement - the quantum dot - 
and the thermal bath 
\begin{equation}
\bar{n} = \frac{\gamma \, \bar{n}_c \,  +\gamma_0 \,  n_B}{\gamma +\gamma_0} \, .
\end{equation}
with $n_B$ the Bose distribution at frequency $\omega_0$ for temperature $T$ of the thermal bath.
Thus ground state cooling $\bar{n} \ll1$ also requires that the electromechanical damping
dominates over the intrinsic damping $ \gamma_0 \,  n_B \ll \gamma \, \bar{n}_c$.
The latter inequality means $\gamma_0/\gamma \ll \bar{n}_c / n_B \ll 1$ 
which is a realistic condition for suspended 
CNT-QD which have huge quality factors $Q_0 = \omega_0 / \gamma_0 \sim 10^6$.

\subsection{Inelastic current}
Finally we discuss the inelastic current associated to the electron-vibration interaction. 
This current results from vibration assisted tunneling processes in which electron hops from a lead to the dot 
exchanging  energy with the oscillator.
Both phonon emission and phonon absorption give a contribution to the inelastic current.
Hence, owing to the discussion of the previous section, one expects a priori the following expression 
for the inelastic current
\begin{equation}
\label{eq:I_in}
I_{in} =\, q^{*} \, \left[ \,  \gamma_{-} \left( \bar{n} + 1 \right) \, + \,  \gamma_{+ } \bar{n} \,  \right] \, .
% \quad \mbox{with} \,\, q^{*}= \, e \, , 2e  \, .
%
\end{equation}
In the section \ref{sec:3} we will discuss the case of a quantum dot coupled to the vibration via the dot's spin, $\hat{F}_d = \hat{s}_x$. 
Then, for the fully spin polarized electrons in the ferromagnetic leads and in the limit of large applied voltage $V$, 
the inelastic current takes indeed the form given by Eq.~(\ref{eq:I_in}) with $q^{*}=e$.
Similarly in section \ref{sec:4} we will discuss the case of a quantum dot coupled to the vibration via the dot's charge,  $\hat{F}_d = \hat{n}_d$. 
Again, in the limit of subgap transport in which the current is mainly determined by  Andreev Reflections 
and in the limit of large applied voltage $V$ (but still $eV \ll \Delta$ with $\Delta$ the superconducting gap), 
the inelastic current reduces to Eq.~(\ref{eq:I_in}) with $q^{*}=2e$ since two electrons are involved 
in the current in order to form a Cooper pair into the superconductor.

%%%%%%%%%%%%%%%%%%%%%%%%%%%%%%%%%%%%%%%%%%%%%%%%%%%%%%
%
%			FIGURE N.2
%
%
%%%%%%%%%%%%%%%%%%%%%%%%%%%%%%%%%%%%%%%%%%%%%%%%%%%%%%
%
%
\begin{figure}[t]
%\resizebox{0.75\columnwidth}{!}{%
%
\resizebox{0.9\columnwidth}{!}{ 
\includegraphics{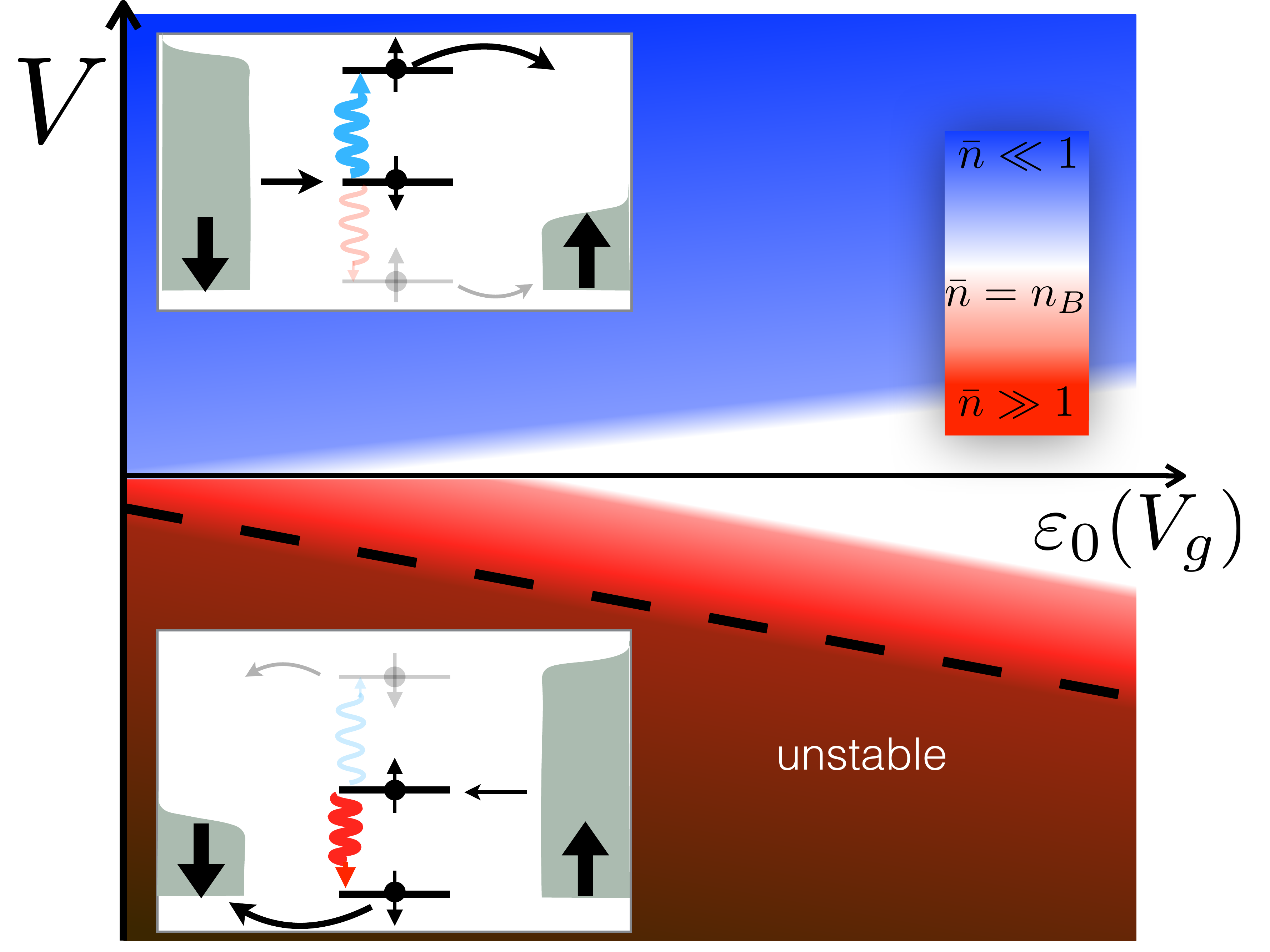}\hspace{2cm}\includegraphics{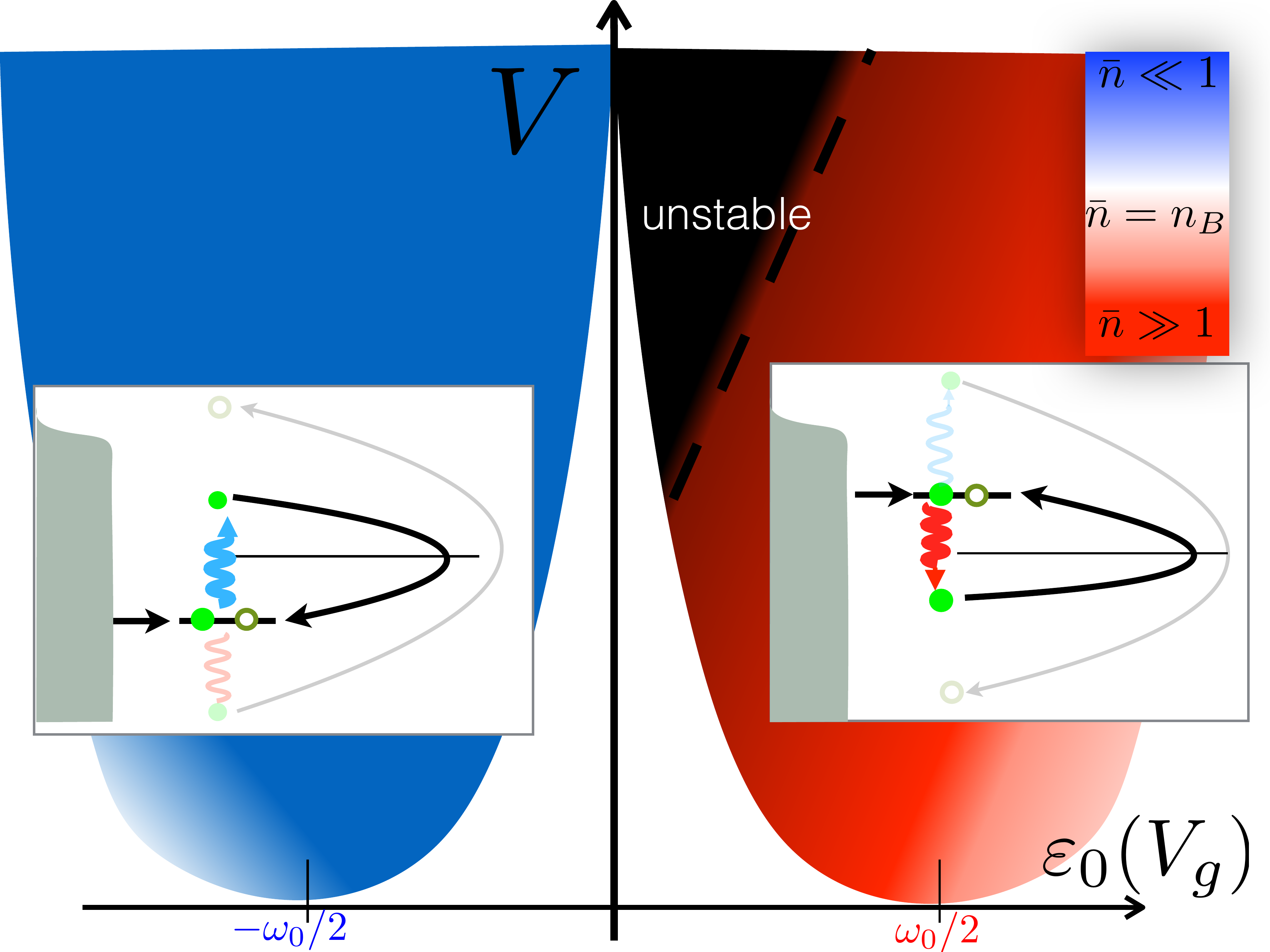}
}
\caption{
Schematic picture of the phonon occupancy  as a function of the bias voltage   
and the average dot's energy  $\varepsilon_0$ (controlled by the gate voltage).  
(a) In the nanomechanical spin-valve,  the figure represents the case of fully polarized leads and the resonance $ \omega_0 = \Delta \varepsilon_z $.
At fixed configuration of the Zeeman splitting in the dot and polarization of the leads,
the  phonon absorption  is enhanced at positive voltage. 
At the negative voltage, the opposite regime occurs. 
(b)  In the system with a superconducting lead,  the Andreev Reflections are mainly associated to the impinging electrons 
at high voltage. The   phonon absorption  is enhanced when the reflected hole appears at the same energy 
of the incoming electron. This is possible if, for example, the electrons enters the dot at the energy $\varepsilon_0 = - \omega_0/2$ 
such that it enters the superconductor at energy $\varepsilon_e = \omega_0/2$. 
Then the hole is reflected at energy $\varepsilon_h = - \omega_0/2$.
The opposite regime occurs when  the electrons enters the dot at the energy $\varepsilon_0 = \omega_0/2$
}
\label{fig:2}
\end{figure}
%
%
%
%%%%%%%%%%%%%%%%%%%%%%%%%%%%%%%%%%%%%%%%%%%%%%%%%%%%%%
 
\section{Spin-vibration interaction and inelastic spin-flip tunneling}
\label{sec:3}

We consider the quantum dot formed by two spin levels with effective Zeeman splitting   
$\varepsilon_z= \varepsilon_{\uparrow}-\varepsilon_{\downarrow}$  and average energy 
$\varepsilon_0= \left( \varepsilon_{\uparrow}+\varepsilon_{\downarrow} \right) / 2$.
To model the spin-valve CNT-QD  embedded between ferromagnetic leads and 
to simplify the discussion, we restrict to the case of fully polarized leads 
such that we can identify $\alpha=l \leftrightarrow  \sigma= \downarrow$ and $\beta=r \leftrightarrow  \sigma= \uparrow$ 
in the Hamiltonian
\begin{equation}
\label{eq:Hamiltonian_H0}
\sum_{\alpha=l,r} \left( \hat{H}_{\alpha}+\hat{H}_{\alpha,t} \right)
= 
\sum_{\sigma=\uparrow,\downarrow}
\sum_k
\left[ 
\varepsilon_{k\sigma}^{\phantom{}} \hat{c}^{\dagger}_{k\sigma}  
\hat{c}_{k\sigma}^{\phantom{}}
+
 t_{\sigma}^{\phantom{}} \hat{c}^{\dagger}_{ k \sigma } 
\hat{d}_{\sigma}^{\phantom{}} 
+
 t_{\sigma}^{*} \
\hat{d}_{\sigma}^{\dagger} 
\hat{c}^{\phantom{}}_{ k \sigma } 
\right]  \, . 
\end{equation}
We also restrict the analysis to the  symmetric contacts such that the tunneling rates 
$\Gamma_{l}^{\uparrow} = \Gamma_{r}^{\downarrow} = \Gamma$.
In a simple picture, for fully polarized leads, the current can flow through the system only if the spin is flipped 
when the electrons pass through the dot. 
This process occurs inelastically with the absorption or the emission of one phonon in weak coupling regime. 
Hence, the system still acts as a nanomechanical spin-valve
in which spin-polarized electrons tunneling through the dot's levels can exchange energy 
with the oscillator by flipping their spins.
At large bias voltage $V$ compared to the other energies (temperature $T$, the tunneling rate $\Gamma$ and the energy spin level 
$\varepsilon_{\sigma}$) the electrons flow pratically from the left to the right, 
as shown in the upper inset of Fig.~\ref{fig:2}(a).
Then the electromechanical damping can be written as $\gamma = \gamma_{+} - \gamma_{-}$
%
% Hereafter in this section we assume this case $eV>0$. 
%
The coefficients $\gamma_{\pm}$ correspond to the rates for vibration-assisted inelastic processes in which a spin flip occurs
for one electron tunneling from the left lead left to right 
accompanied by the absorption ($s =+$) or emission ($s=-$) of an vibrational energy quantum $\omega_0$.  
\begin{equation}
\gamma_{\pm}
=
\lambda^2  \Gamma^2  \,  \int \!\!\!  d\omega  \,\,
T_{\pm} (\omega) \,\, 
 f_{l}(\omega)\left[1\mathord- f_{r}( \omega \pm \omega_0)\right] 
\simeq  
\lambda^2  \Gamma^2   \,  \int \!\!\!  d\omega  \,\, T_{\pm} (\omega) \,\, \, , 
\label{eq:gamma_spin} 
\end{equation}
where $f_{l,r}(\omega) = 1 / \{ 1 + \exp[( \omega - \mu_{l,r})/T]  \}  $ are the Fermi functions at the left and right lead,
$\mu_l-\mu_r=eV$ (approximately $f_{l}(\omega) \simeq 1$ and $f_r(\omega) \simeq 0$  for high voltage), 
whereas  the transmission functions  read
\begin{equation}
\label{eq:T_spin}
T_{\pm}(\omega) = \frac{1}{\pi} 
\frac{ \Gamma^2 }{\left[ \Gamma^2+(\omega-\varepsilon_{\downarrow})^2\right] } 
\frac{ \Gamma^2}{\left[ \Gamma^2+(\omega+s\omega_0 -\varepsilon_{\uparrow})^2\right] }   \, .
\end{equation}
The Eq.~(\ref{eq:gamma_spin}) can be evaluated analitically.
We report the result for the resonace case $\varepsilon_{\uparrow} - \varepsilon_{\downarrow} = \omega_0$ which reads
\begin{equation}
\label{eq:gamma_spin_resonance}
\gamma_{+} = \frac{\lambda^2}{2\Gamma}      \, , \qquad 
\gamma_{-} =    
\gamma_{+} \, \frac{ \Gamma^2 }{\Gamma^2 +  \omega_0^2  }  \simeq 
 \gamma_{+} \,  {\left( \frac{ \Gamma }{  \omega_0 } \right)}^2 \, ,
\end{equation}
%
%
%\frac{\pi}{2}  \left( \frac{\lambda^2 \Gamma^4}{\Gamma^2 +  \omega_0^2  } \right)
%
%
from which we extract the minimum values of the phonon occupations that can be achieved, 
namely $n_{min} \simeq \gamma_{-}/\gamma_{+} = {( \Gamma/\omega)}^2$.
The situation changes at negative voltage where we have a region of increase of the phonon occupation 
$n \gg 1$ for  $\gamma_{+} \gtrsim \gamma_{-}$ and an instability region when $\gamma_{+} < \gamma_{-}$.
These two regions are beyond the validity of the  perturbative approach and the phase diagram represents 
only a qualitative description.

The results of  Eqs.~(\ref{eq:gamma_spin},\ref{eq:T_spin},\ref{eq:gamma_spin_resonance}) 
enlighten the ultimate mechanism for the cooling.
The two Lorentzian functions in the integral  of Eq.~(\ref{eq:gamma_spin})  
completely overlaps for the case of the absorption rate  $s = +$ in the cooling region. 
In other words, the inelastic spin-flip occurs through the two peaked spin levels of the dot's density of states. 
Oppositely,  in the case of emission $s=-$,  the two Lorentzian functions in the integral  of Eq.~(\ref{eq:gamma_spin}) 
are well separated: phonon emission is still possible but arises through only one peak associated to the spin down 
whereas the passage through the spin up can be seen as a cotunneling process whose amplitude scales as 
$\sim  \Gamma/\omega_0 \ll 1$.

Finally we discuss the behavior of the inelastic current in the limit case when the oscillator is strongly affect by the 
quantum dots and the steady state phonon occupation saturates to $\bar{n} \simeq \bar{n}_c$.
In this regime the current clearly reflects the behavior of the phonon occupancy.
At large positive  voltage, in the cooling regime, we have $\bar{n}_c \ll 1$ and $\gamma_{-} \ll \gamma_{+}$ 
\begin{equation}
I_{in}^{eV>0}   \simeq  e \gamma_{-}  =   I_0 \left( \frac{\lambda^2}{2 \omega_0^2}   \right)
\end{equation}
with $I_0 = e \Gamma$.
In the cooling regime the inelastic current is strongly suppressed with respect to the elastic current of order $I_0$.
At negative  voltage with $\gamma_{+} \gtrsim \gamma_{-}$  
we have $\bar{n}_c \geq 1$   such that we can approximate 
\begin{equation}
I_{in}^{eV<0}   \simeq  \, e \, \gamma \, \bar{n}_c   \, .
\end{equation}
Since the phonon occupation scales as $\bar{n}_c  \sim \gamma_{-}/(\gamma_{+}-\gamma_{-})$, it increases 
indefinitively as long $\gamma_{+} \rightarrow \gamma_{-}$ until the instability $\gamma_{+}-\gamma_{-}<0$.
In summary, a strong asymmetry
emerges in the inelastic current that reflects the behavior of the nonequilibrium 
phonon occupation $\bar{n}_c$.

\section{Charge-vibration interaction and inelastic Andreev Reflection}
\label{sec:4}
The model Hamiltonian for a superconductor/ normal metal quantum dot  is formed by the Fermi reservoir 
$H_{l} = H_{N} = \sum_k \varepsilon_{k\sigma}^{\phantom{}} \hat{c}^{\dagger}_{k\sigma}  
\hat{c}_{k\sigma}^{\phantom{}}$  and the BCS Hamiltonian
$H_{r} = H_{S} 
=  \sum_k \left[ \epsilon_{k\sigma}^{\phantom{}} \hat{f}^{\dagger}_{k\sigma}  \hat{f}_{k\sigma}^{\phantom{}}
+\Delta \left( \hat{f}^{\dagger}_{k\uparrow}  \hat{f}_{-k\downarrow}^{\dagger} +\mbox{h.c.} \right)
\right]$ and the tunneling Hamiltonian 
\begin{equation}
\label{eq:Hamiltonian_H0}
\hat{H}_{N,t} +\hat{H}_{S,t} 
=
 \sum_{\sigma=\uparrow,\downarrow}
\sum_k  \left( t_N \hat{c}^{\dagger}_{ k \sigma } 
\hat{d}_{\sigma}^{\phantom{}} + t_S  \hat{f}^{\dagger}_{ k \sigma } 
\hat{d}_{\sigma}^{\phantom{}} 
\, + \,  \mbox{h.c.} \right)
  \, . 
\end{equation}
In the strong subgap regime, defined by the condition that the gap $\Delta$ is the largest energy scale in
the problem, the charge transport through the quantum dot occurs via Andreev Reflection (AR)
whose transmission amplitude is independent of $\Delta$.
In this case, the relevant quantity are the tunnelling rates from the normal lead to the dot $\Gamma_N$
and the  tunnelling rates from the superconductor  to the dot $\Gamma_S$.

An electron at energy much lower than the energy gap and and tunnelling on the quantum dot from the normal metal 
can be either simply inelastic reflected either inelastic reflected as hole (AR).
Thus the electromechanical damping is associated to  these two inelastic process   
$\gamma = \gamma_{NR} + \gamma_{AR}$.
However the normal reflection (NR) can drive the oscillator only to the thermal equilibrium:  in these processes, 
the oscillator is affected by only one fermionic reservoir  at unique temperature $T$.
Hence inelastic normal reflection forms an additional mechanism of normal damping and $\gamma_{NR}$ adds to the 
intrinsic damping $\gamma_0$. 
By contrast, the inelastic ARs can drive the resonator towards a nonequilibrium steady state.
From now on we focus on  the inelastic AR  processes.
%
%	and neglect the norma reflections which corresponds to a good approximation 
%	in the case of strongly asymmetric tunneling coupling $\Gamma_S\gg\Gamma_N$,  
%
%
Setting the chemical potential of the superconductor  $\mu= 0$, 
we consider the high voltage
limit (but still 
$eV \ll \Delta $
)  in which the current can be described as given by impinging
electrons that are reflected as holes.
Then the emission/absorption rates reads
\begin{equation}
\gamma_{\pm}
=
\lambda^2   \, \Gamma_N^2 \int \!\!\!  d\omega  \,\,
T_{\pm} (\omega) \,\, 
 f(\omega)\left[1\mathord- \bar{f} ( \omega \pm \omega_0)\right] 
\simeq  
\lambda^2   \, \Gamma_N^2 \,  \int \!\!\!  d\omega  \,\, T_{\pm} (\omega) \,\, \, , 
\label{eq:gamma_AR} 
\end{equation}
where, beyond the Fermi occupation function $f(\omega) = 1 / \{ 1 + \exp[( \omega - eV)/T]  \} \simeq 1$,  
we have introduce the  occupation function for the holes
$\bar{f}(\omega) = 1 / \{ 1 + \exp[( \omega + eV)/T]  \} \simeq 0$, and the transmission function 
\begin{equation}
\label{eq:T_AR}
T_{\pm}(\omega) =  \frac{1}{4\pi}
{\left| 
G_e\left( \omega \right)  F^*\left( \omega + s \omega_0 \right) -   F\left( \omega  \right) G^*_h\left(  \omega + s \omega_0 \right) 
\right|}^2
  \, ,
\end{equation}
with the Green functions defined as 
\begin{eqnarray}
G_{e/h}(\omega) &=& \frac{ \omega \pm \varepsilon_0 + i \Gamma_N }{ \left(  \omega + \varepsilon_0 + i \Gamma_N  \right)  \left( \omega - \varepsilon_0 + i \Gamma_N \right) - \Gamma_S^2  }		\label{eq:G_F_1} \, ,\\
F (\omega)  &=& \frac{ \Gamma_S }{ \left(  \omega + \varepsilon_0 + i \Gamma_N  \right)  \left( \omega - \varepsilon_0 + i \Gamma_N \right) - \Gamma_S^2  } \, .
\label{eq:G_F_2} 
\end{eqnarray}
The square modulus of the anomalours Green function $F (\omega)$ plays the role of transmission function in the formula for the elastic current 
associated to ARs through the dot (for instance, an incoming electron at energy $\omega$).
Thus one can regard it as the effective amplitude for the AR.
The other two   functions $G_{e/h}(\omega)$ are the electron and hole Green functions of the dot in tunneling contact with  the superconductor 
and they play the role of transmission function in the tunneling in the normal case 
(for instance,  an incoming electron at energy $\omega$). 
In other words, the transmission function consists of a coherent sum of two amplitudes that are associated 
to the two possible paths in which the phonon is emitted or absorbed before or after an AR. 
The integral of the transmission function in the last term of Eq.~(\ref{eq:gamma_AR}) can be done analitically and 
we obtain
\begin{eqnarray}
\gamma_{\pm}(\varepsilon_0)&=& \, 
\lambda^2\,  \Gamma_S^2 \, \Gamma_N \, 
\frac{ \left( E_A^2 + \frac{\omega_0^2}{4} + 5\Gamma_N^2\right) }{
\left(E_A^2+\Gamma_N^2\right) \left( \frac{\omega_0^2}{4} + \Gamma_N^2\right)
} \nonumber \\
&\times&  \frac{  {\left(\pm \omega_0/2 - \varepsilon_0 \right)}^2  + \Gamma_N^2}{ 
\left[ {(\pm\omega_0/2)- E_A}^2 + \Gamma_N^2\right]
\left[ {(\pm \omega_0/2 + E_A)}^2 + \Gamma_N^2\right]
 } 
\label{eq:gamma_AR}
\end{eqnarray}
with $E_A = \sqrt{\varepsilon_0^2 + \Gamma_S^2}$ and $s=+/-$ for the absorption and emission.
Remarkably, the rate for phonon emission is strongly suppressed at $\varepsilon_0 = - \omega_0/2$ such that 
the resonator approaches the ground state with minimum phonon occupation 
$n_{min} = \gamma_{-}/\gamma_{+} \simeq {(\Gamma_N / \omega_0 )}^2 $, see Fig.~\ref{fig:2}(b). 
At the symmetric point $\varepsilon_0 =  \omega_0/2$,   
the  rate for phonon absorption is strongly reduced and we are in the full region of instability 
$\gamma_{+} \ll \gamma_{-}$, see Fig.~\ref{fig:2}(b). 
The final result Eq. ~(\ref{eq:gamma_AR}) is a consequence of the form of the
transmission function Eq.~(\ref{eq:T_AR}): ground state cooling is achieved due to the destructive
interference of the two amplitudes associated to the charge transmission with phonon
emission.

As for the previous system, we discuss the behavior of the inelastic current in the limit case when the oscillator is strongly affect by the 
quantum dots and the steady state phonon occupation saturates to $\bar{n} \simeq \bar{n}_c$.
In contrast to the previous system of section \ref{sec:3}, the current has  a sharp dependence 
on the dot's energy levels $\varepsilon_0$.
We give an example assuming the case 
$\Gamma_S \ll |\varepsilon_0|,\omega_0$ and $|\varepsilon_0|  \approx \omega_0/2$.
In the cooling regime, with $\varepsilon_0 < 0$,  we have $\bar{n}_c \ll 1$ and $\gamma_{-} \ll \gamma_{+}$ 
and we can approximate $I_{in} \simeq \, 2 \, e \, \gamma_{-}$.  
For  $\varepsilon_0 \approx -\omega_0/2$, the inelastic current shows a peak 
\begin{equation}
I_{in} \left(\varepsilon_0 \approx -\omega_0/2 \right) = I_{in}^{-}
   \,\,  \simeq \,\,  2 e
 \frac{ 8 \lambda^2 \Gamma_S^2   \Gamma_N^3}{  
\omega_0^4  \left[{(\varepsilon_0  + \omega_0/2)}^2 + \Gamma_N^2\right]
 } \, .
\end{equation}
In the regime $\varepsilon_0 > 0$  with $\gamma_{+} \gtrsim \gamma_{-}$  
we have $\bar{n}_c \geq 1$   such that we can approximate $I_{in} \simeq 2 e \gamma  \bar{n}_c$.
Close to  $\varepsilon_0 \approx \omega_0/2$ (but far away the instability region),  we can approximate the peak of the inelastic current to  
\begin{equation}
I_{in} \left(\varepsilon_0 \approx \omega_0/2 \right)= I_{in}^{+}
\simeq 
 2 e
 \frac{ 8 \lambda^2 \Gamma_S^2   \Gamma_N^3}{  
\omega_0^4  \left[{(\varepsilon_0  - \omega_0/2)}^2 + \Gamma_N^2\right]
 } 
 \,
 \frac{\bar{n}_c(\omega_0/2)}{  n_{min}  }   \, .
\end{equation}
In such nonequilibrium regime of the oscillator, we conclude that 
the peak around $\varepsilon_0 \approx \omega_0/2$ will be higher than the peark at  $\varepsilon_0 \approx -\omega_0/2$
since the first one is enhanced by the phonon occupation $\bar{n}_c(\omega_0/2) \geq 1$  
and by the factor $n_{min}= {( \Gamma_0/\omega_0)}^2$. 
Furthermore, since the phonon occupation scales as $\bar{n}_c \sim \gamma_{-}/(\gamma_{+}-\gamma_{-})$, it increases 
indefinitively as long $\gamma_{+} \rightarrow \gamma_{-}$ until the instability occurs $\gamma_{+}-\gamma_{-}<0$.

To summarize, in the case of a quantum dot with charge-vibration interaction inducing inelastic ARs,
the effect of the coupling with the resonator appears in the sub-gap transport as sharp, vibrational side band peaks
which are not broadened by the temperature of the normal leads.
Moreover a  strong asymmetry of the two peaks points out clearly the nonequilibrium   state of the oscillator.

\section{Conclusions}
\label{sec:5}
To conclude, we have presented two theoretical proposals for controlling the nonequiibrium steady 
state of nanomechanical resonators integrating quantum dots.
One of the main results is that ground state cooling of the resonator can be realistically
achieved using spin-polarised current \cite{Stadler:2014,Stadler:2015}
or a superconducting contact \cite{Stadler:2014,Stadler:2015}. 
For the two differerent proposals, we have also shown how the nonequilibrium states of the
resonator can be readily detected by simple measurements of the dc current.
Finally, we remark that the on-site charging energy, that we have neglected in our analysis,  does not 
break qualitatively our findings.
For the case of the spin-vibration interaction,  correlation effects  associated to the double occupation 
eventually set the charge flow but do not prevent the occurrence of inelastic spin-flip tunneling \cite{Stadler:2015}. 
For the charge-vibration interaction, the Andreev Reflections rely on superconducting correlations in the quantum dot.
% of  coherent superposition of double occupied state and empty state.
%
Indeed, in the superconducting gap limit $\Delta \rightarrow \infty$ and  high voltage limit (but still $eV \ll \Delta$), 
it is still  possible  to  establish  a  BCS-like state in the quantum dot even in the presence of strong Coulomb repulsion 
when the tunneling coupling between the superconductor and the quantum  dot  is larger than  the tunneling coupling  
with the normal  lead \cite{Braggio:2011}. 

\section{Acknowledgments}
\label{sec:6}
We thanks Pascal Stadler for useful discussions. This work was supported by the
Excellence Initiative through the Zukunftskolleg and by the DFG through the SFB 767.

%%%%%%%%%%%%%%%%%%%%%%%%%%%%%%%%%%%%%%%%%%%%%%%%%%%%%%%%%%%%%%%%%%%%%%%%%
%
%
%
%\bibliographystyle{acm}
%\bibliography{references}
%
%
%
%%%%%%%%%%%%%%%%%%%%%%%%%%%%%%%%%%%%%%%%%%%%%%%%%%%%%%%%%%%%%%%%%%%%%%%%%

%
\end{document}